\documentclass[aps,pra,reprint,amsmath,amssymb,superscriptaddress,nobibnotes, longbibliography]{revtex4-1}
\usepackage{graphicx,SIunits}
\usepackage{bm}     
\usepackage{hyperref} 
\usepackage{dsfont}    
\usepackage{color}

\usepackage{amsmath}
\usepackage{amsthm}
\usepackage{amssymb}
\usepackage{tikz}

\usetikzlibrary{decorations.pathreplacing}
\usepackage{braket}
\usepackage{dsfont}

\usepackage{color}
\usepackage{lineno}

\begin{document}

\title{Demonstration of subspace-search variational quantum eigensolver using photonic orbital angular momentum qudits}

\author{Kang-Min Hu}
\email{These authors contributed equally}
\affiliation{Center for Quantum Technology, Korea Institute of Science and Technology (KIST), Seoul, 02792, Korea}
\affiliation{Division of Quantum Information, KIST School, Korea University of Science and Technology, Seoul, 02792, Korea}

\author{Myung-Hyun Sohn}
\email{These authors contributed equally}
\affiliation{Center for Quantum Technology, Korea Institute of Science and Technology (KIST), Seoul, 02792, Korea}
\affiliation{Department of Applied Physics, Kyung Hee University, Yongin, 17104, Korea}

\author{Yong-Su Kim}
\affiliation{Center for Quantum Technology, Korea Institute of Science and Technology (KIST), Seoul, 02792, Korea}
\affiliation{Division of Quantum Information, KIST School, Korea University of Science and Technology, Seoul, 02792, Korea}

\author{Kwang Jo Lee}
\affiliation{Department of Applied Physics, Kyung Hee University, Yongin, 17104, Korea}

\author{Hyun Woo Kim}
\affiliation{Center for Quantum Technology, Korea Institute of Science and Technology (KIST), Seoul, 02792, Korea}
\affiliation{Department of Chemistry, Gwangju Institute of Science and Technology, Gwangju, 61005, Korea}

\author{Hyang-Tag Lim}
\email{Corresponding author: forestht@gmail.com}
\affiliation{Center for Quantum Technology, Korea Institute of Science and Technology (KIST), Seoul, 02792, Korea}
\affiliation{Division of Quantum Information, KIST School, Korea University of Science and Technology, Seoul, 02792, Korea}

\date{\today} 

\begin{abstract}
Accurate excited state energies are essential for interpreting fluorescence and photochemical processes, yet variational quantum eigensolvers (VQEs) on noisy intermediate-scale quantum (NISQ) hardware have largely been confined to ground states. Subspace-search VQE (SSVQE) offers a NISQ-friendly route to excited states without ancilla-based overlap measurements, however it has lacked an experimental demonstration since it requires the coherent preparation and manipulation of multiple orthogonal basis states under a shared variational unitary. Here we realize SSVQE on a photonic platform by encoding orthogonal qudit states in the orbital angular momentum of single photons and optimizing a shared variational circuit. We estimate excited state energies for a four-dimensional $\rm{HeH^+}$ molecule and an eight-dimensional Heisenberg chain, thereby demonstrating scalable excited state eigensolving in a high-dimensional Hilbert space. Our results establish photonic qudits as a resource-efficient pathway to excited state quantum simulation.
\end{abstract}

\maketitle

\section*{Introduction}

The computation of molecular Hamiltonians to explain the properties of matter constitutes a fundamental problem in the physical and chemical sciences. Excited states play a crucial role in a wide range of phenomena, including fluorescence~\cite{itoh2012}, chemical reactions~\cite{lischka2018}, and hydrogen bonding~\cite{zhao2012}. Although classical methods such as density functional theory are widely used and usually provide accurate ground state energies, they do not always yield reliable descriptions of excited states~\cite{dreuw2004,levine2007}. These limitations have motivated quantum computational approaches for estimating excited states. However, many such methods rely on controlled-unitary operations, as required for the iterative phase estimation algorithm (IPEA), thereby presenting significant experimental challenges--particularly in high-dimensional systems~\cite{lanyon2010,santagati2018}.

In the noisy intermediate-scale quantum (NISQ) era, the variational quantum eigensolver (VQE) has emerged as a prominent quantum--classical hybrid algorithm~\cite{preskill2018,hu2025}, primarily employed for estimating the ground state energy of a given Hamiltonian. VQE prepares a parametrized quantum state (ansatz) and iteratively optimizes its parameters to minimize the expectation value of the Hamiltonian~\cite{peruzzo2014,tilly2022}. Owing to its potential applicability to complex Hamiltonians, VQE has been actively investigated across a range of hardware platforms, including superconducting devices~\cite{google2020,kandala2017}, trapped ions~\cite{hempel2018}, and photonic systems~\cite{peruzzo2014,santagati2018,kim2024,maring2024,lee2022,lee2024,baldazzi2025,hoch2025}. Nevertheless, the standard formulation of VQE is fundamentally limited to ground state estimation.

To overcome this limitation and access excited states, a variety of extensions to VQE have been proposed~\cite{bidart2025}, including quantum subspace expansion~\cite{mcclean2017,colless2018,yoshioka2022},  equation-of-motion approaches~\cite{ganzhorn2019, ollitrault2020}, variational quantum deflation~\cite{higgot2019}, constraint-based VQE~\cite{ilya2018,kuroiwa2021,gocho2023,parrish2019}, discriminative VQE~\cite{tilly2020}, and variational imaginary-time evolution~\cite{jones2019}. While these methods provide systematic routes to excited state estimation, they typically require classical diagonalization or the solution of large linear systems, and/or rely on ancilla-assisted overlap estimation, controlled operations, and penalty-type hyperparameters. These requirements generally lead to deeper circuits and increased optimization overhead. In contrast, subspace-search VQE (SSVQE) can, in principle, recover the exact spectrum within a chosen subspace under ideal unitaries while avoiding explicit inner-product estimation, making it particularly well suited to NISQ devices~\cite{nakanishi2019}. Despite this promise, its experimental realization has remained elusive since SSVQE requires high-fidelity preparation of multiple mutually orthogonal input states and the reproducible application of a common parametrized unitary across them.

In this work, we demonstrate an experimental implementation of SSVQE on a photonic platform by exploiting high-dimensional qudit encoding in the orbital angular momentum (OAM) of single photons. Rather than generating the required orthogonal inputs through deep gate sequences, we directly prepare a set of mutually orthogonal OAM qudit states using holographic patterns on a spatial light modulator (SLM), which is advantageous for high-dimensional scalability~\cite{fickler2012,fickler2016,krenn2014}. We then apply a shared variational circuit to all input states, and process measurement outcomes classically to evaluate the SSVQE cost function and update the variational parameters until convergence. We apply this approach to estimate excited state energies of a four-dimensional $\rm{HeH}^+$ molecule and an eight-dimensional Heisenberg chain, thereby validating SSVQE in a high-dimensional photonic platform. Our results establish photonic OAM qudits as a scalable and resource-efficient architecture for excited state eigensolving in high-dimensional Hilbert spaces.

\section*{Results}
\subsection*{SSVQE protocol}
For a Hamiltonian \(H\) with eigenvalues $\{E_j\}_{j=0}^{d-1}$ ordered as $E_0\le E_1\le\cdots$, any normalized trial state $|\psi(\vec{\alpha})\rangle$ satisfies the variational bound
\begin{equation}
\langle{\psi(\vec{\alpha})} |H|\psi(\vec{\alpha})\rangle \ge E_0,
\label{eq1}
\end{equation}
which forms the basis of the VQE~\cite{peruzzo2014}. In VQE, the energy is evaluated as the expectation value of $H$, and the parameters $\vec{\alpha}$ are optimized to minimize Eq. (\ref{eq1}). Using the Jordan-Wigner transformation~\cite{jordan1928}, a multi-qubit Hamiltonian can be decomposed into a weighted sum of Pauli strings,

\begin{equation}
    H =\sum_{m} {W_m P_m},
\end{equation}
where $W_m \in \mathbb{R}$ and $P_m \in \{I,X,Y,Z\}^{\otimes n_q}$, where $n_q$ is the number of qubits. The energy expectation value can then be obtained from Pauli-string measurements as
\begin{equation}
	\langle H \rangle =\sum_{m} {W_m \langle P_m \rangle}.
\label{eq3}
\end{equation}

\begin{figure}[h]%
    \centering
    \includegraphics[width=1\columnwidth]{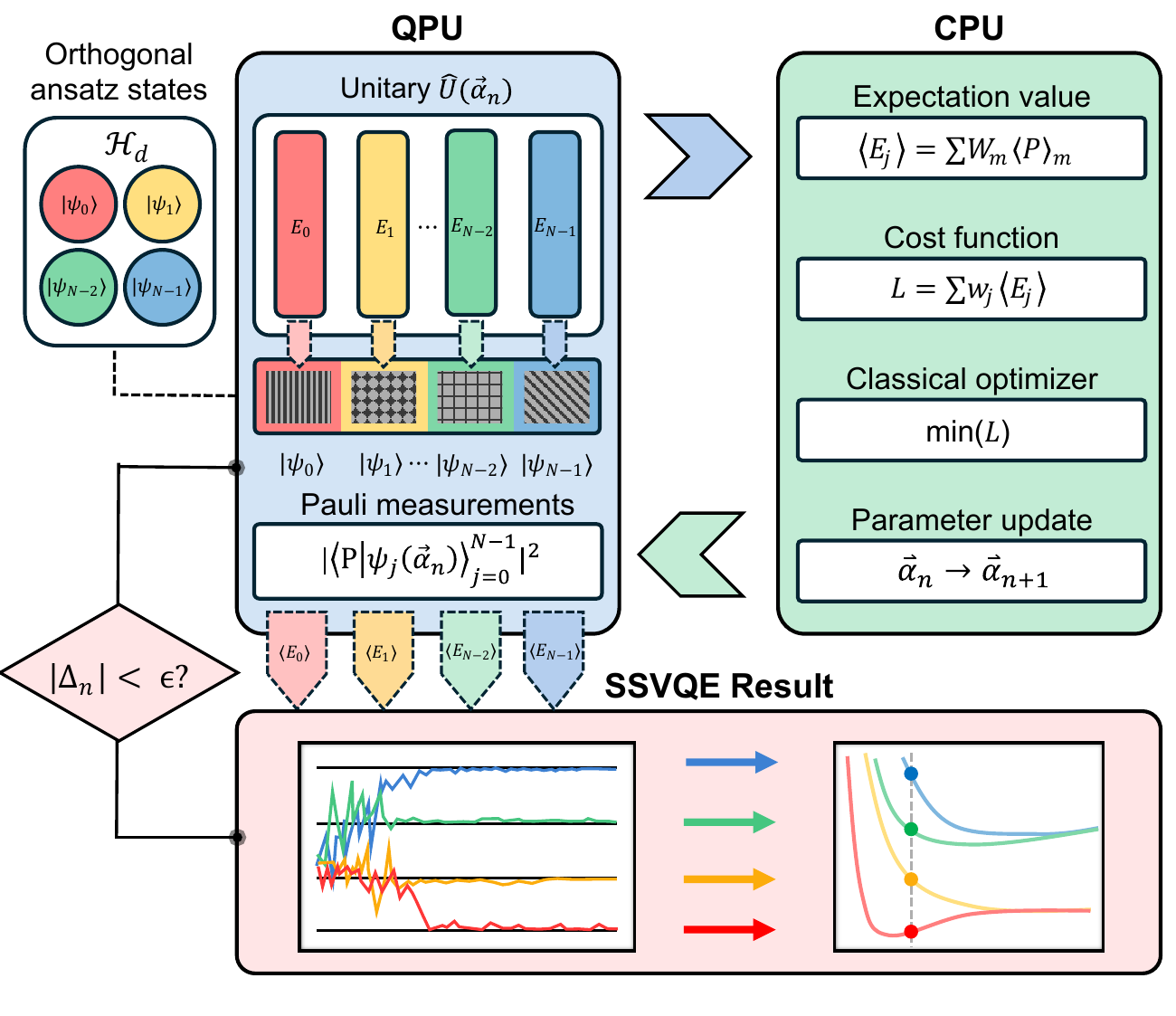}
    \caption{\textbf{OAM qudit--based SSVQE procedure.} At iteration $n$, a set of $N$ mutually orthogonal variational states $\{|\psi_j(\vec{\alpha}_n)\rangle\}_{j=0}^{N-1}$ is prepared by displaying the corresponding holograms on a SLM. Each state is measured on the photonic QPU (quantum processing unit) using a set of Pauli-string observables (implemented with a second SLM), from which the Hamiltonian expectation values are estimated. The CPU (classical processing unit) evaluates the SSVQE cost function $L$ and a classical optimizer (COBYLA) updates the parameters $\vec{\alpha}_{n+1}$ to minimize $L$. The procedure is repeated until convergence, $|\Delta_n|=|\vec{\alpha}_{n+1}-\vec{\alpha}_n|<\epsilon$, yielding the optimized energies $\langle E_0\rangle_{\min}, \langle E_1\rangle_{\min}, \ldots, \langle E_{N-1}\rangle_{\min}$ simultaneously. }
    \label{fig1}
\end{figure}

\begin{figure*}[htb]

\includegraphics[width=2\columnwidth]{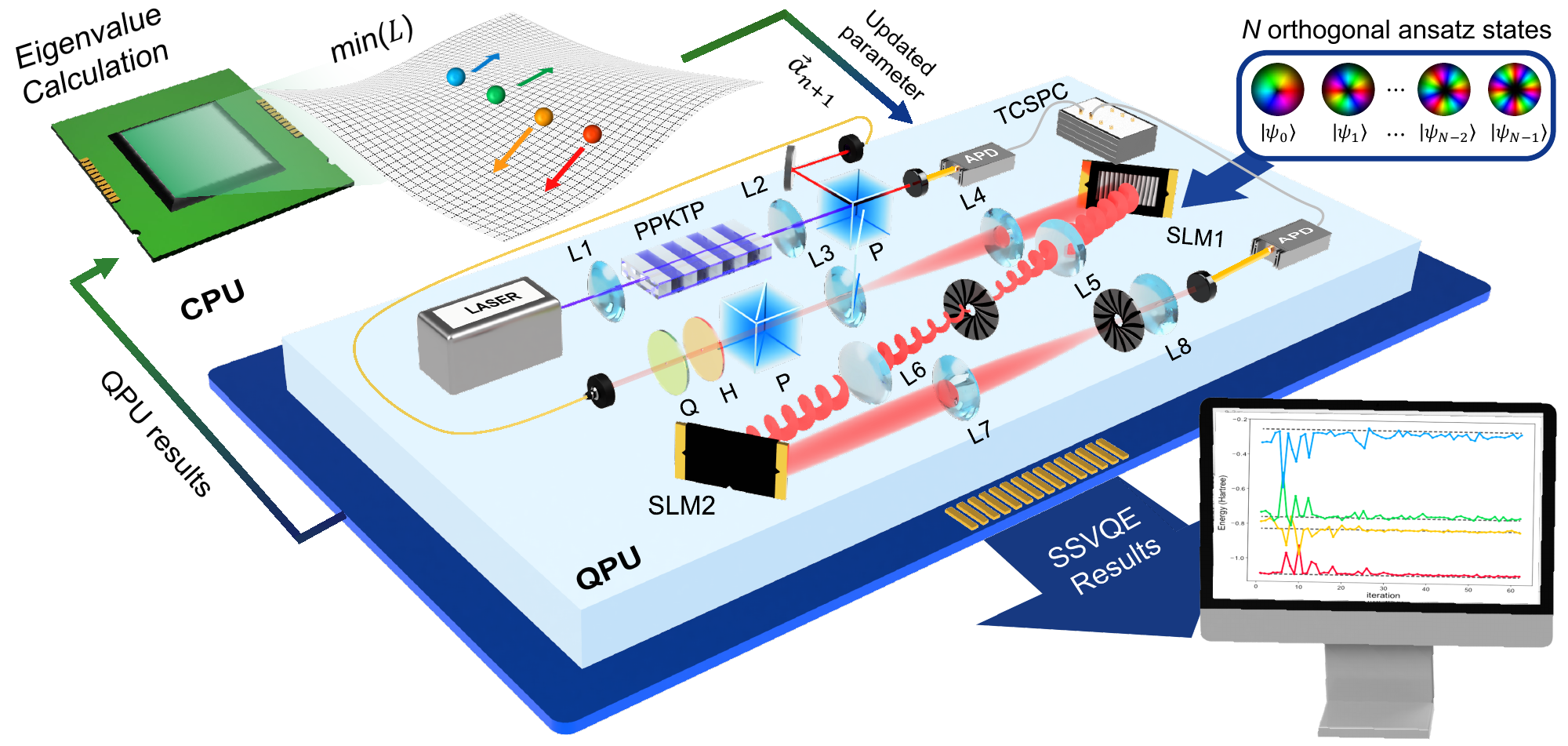}
\caption{\textbf{Experimental setup for SSVQE.} Photon pairs are generated via type-II SPDC; the idler photon is detected by APD1 to herald the signal photon, which is sent to the QPU for SSVQE. In the QPU, the signal photon is prepared in the $j$th OAM ansatz qudit state $| \psi_j (\vec{\alpha}_{n}) \rangle $ by displaying the corresponding hologram on SLM1. The diffracted beam is spatially filtered (iris) and analyzed using measurement holograms on SLM2 to perform Pauli-string measurements. The measurement outcomes are processed by the CPU to evaluate the cost function $L$ and update the variational parameters $\vec{\alpha}$ via a classical optimizer. (SPDC, spontaneous parametric down-conversion; H, Half-wave plate; Q, Quarter-wave plate;  P, Polarizing beam splitter; SLM, Spatial light modulator; APD, Avalanche photodiode; TCSPC, Time-correlated single photon counter.)}
\label{fig2}
\end{figure*}

SSVQE extends this framework to simultaneously target excited states by exploiting orthogonal input states and a shared variational unitary~\cite{nakanishi2019}. Specifically, SSVQE considers a set of $N$ mutually orthogonal input states $|\phi_j\rangle_{j=0}^{N-1}$ and a parametrized ansatz circuit $U(\vec{\alpha})$. For each input state, the corresponding energy estimate is

\begin{equation}
\langle E_j (\vec{\alpha}) \rangle = \langle{\phi_j}|U^\dagger(\vec{\alpha})HU(\vec{\alpha})|{\phi_j}\rangle,
\end{equation}
which can be evaluated using the same Pauli-string decomposition as in Eq. (\ref{eq3}). The SSVQE cost function is defined as a weighted sum,
\begin{equation}
L(\vec{\alpha})=\sum_{j=0}^{N-1} w_j\,\big\langle E_j (\vec{\alpha})\rangle,
\qquad
w_0> w_1> \cdots > w_{N-1},
\label{eq5}
\end{equation}
and minimizing $L(\vec{\alpha})$ drives $\langle E_j (\vec{\alpha})\rangle$ toward the $j$th eigenenergy $E_j$ within the  target subspace.

A key advantage of SSVQE is that orthogonality is enforced at the input and preserved under the shared unitary $U(\vec{\alpha})$, eliminating the need for ancilla-assisted overlap-estimation circuits that are commonly required in deflation-based~\cite{higgot2019} or penalty-based methods~\cite{ilya2018,kuroiwa2021,gocho2023,parrish2019}. Moreover, SSVQE does not require additional measurement settings beyond those used in standard VQE, allowing existing measurement-grouping strategies and error-mitigation techniques to be directly applied~\cite{lee2022,lee2024}.

To realize SSVQE experimentally on a photonic platform, we implement the variational states directly in a high-dimensional OAM qudit space. Instead of preparing $N$ orthogonal inputs and physically implementing a full multi-mode unitary transformation $U(\vec\alpha_n)$ on each input state, we directly prepare the $N$ orthogonal output states $|\psi_j(\vec{\alpha_n})\rangle_{j=0}^{N-1} = U(\vec{\alpha_n})|\phi_j\rangle_{j=0}^{N-1}$ as OAM qudit states. As illustrated in Fig.~\ref{fig1}, the classical processing unit (CPU) calculates the target output-state coefficients from the variational parameters, converts them into the OAM basis, and generates the corresponding phase-amplitude holograms. The hologram corresponding to $|\psi_j(\vec{\alpha}_n)\rangle$ is generated from the $j$th column of $U(\vec{\alpha}_n)$ at iteration $n$ and displayed on the SLM. Thus, rather than realizing the entire unitary matrix, our implementation synthesizes the relevant output states, equivalently the columns of the effective variational unitary required for the SSVQE cost-function evaluation. Each state is then physically prepared and measured on the photonic QPU with the same set of Pauli-string observables to estimate $\langle E_j (\vec{\alpha}_n)\rangle$ from photon-counting statistics, after which the CPU evaluates $L(\vec{\alpha}_n)$ and updates the parameters $\vec{\alpha}_{n+1}$ to minimize Eq.~(\ref{eq5}). The procedure is repeated until convergence, $|\vec{\Delta}_n|= | \vec{\alpha}_{n+1} - \vec{\alpha}_n| < \epsilon$, where $\epsilon$ is the convergence tolerance, yielding the estimated energies for the $N$ lowest-lying eigenstates within the chosen subspace.

\begin{figure*}[!t]
    \centering
    \includegraphics[width=2\columnwidth]{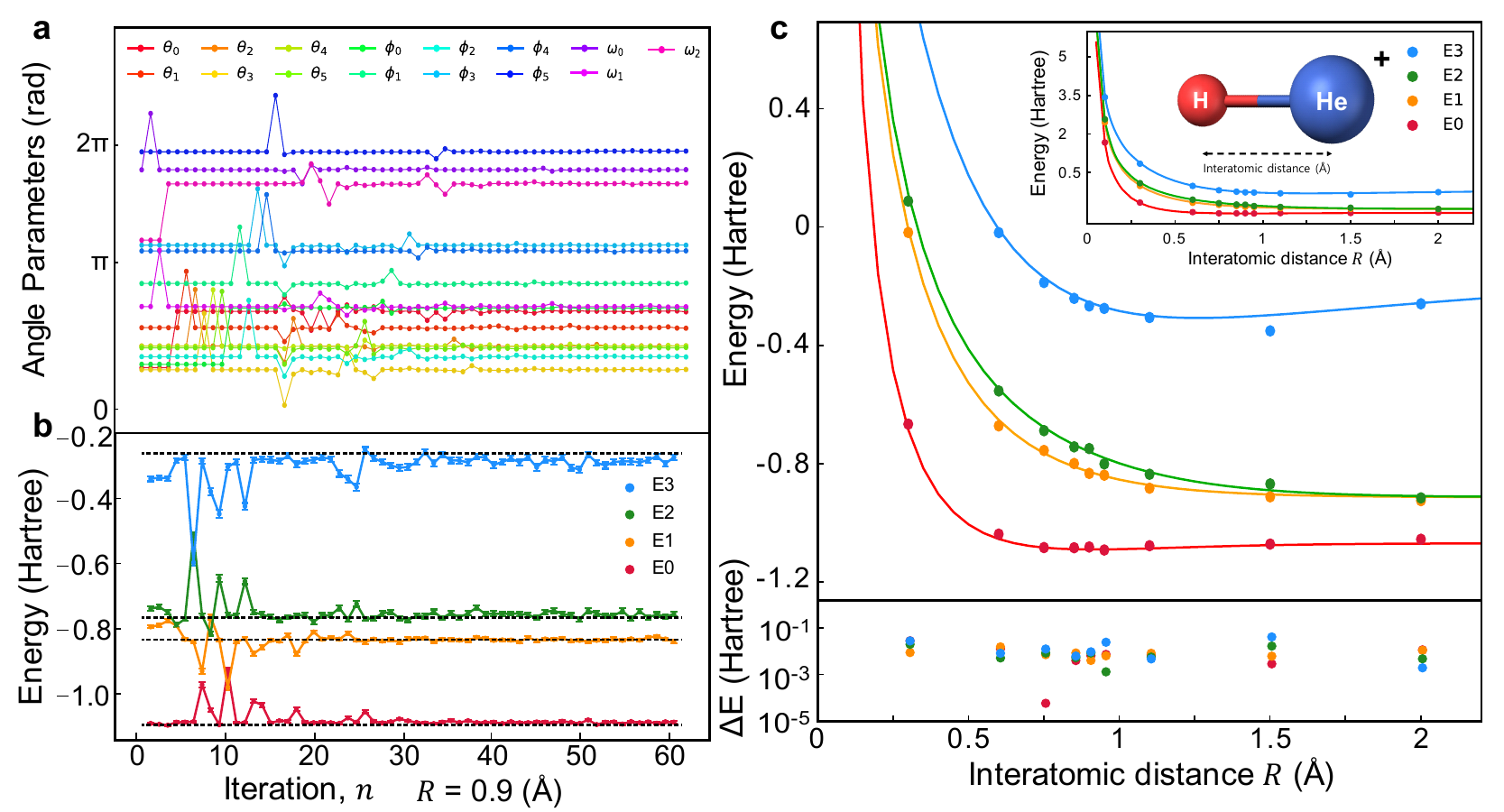}
    \caption{\textbf{Experimental SSVQE results for HeH$^{+}$.} \textbf{a,} Iterative evolution of the 15 variational angle parameters that control the four OAM-qudit ansatz states at $R=0.90~\text{\AA}$ (equilibrium bond length). The parameter updates gradually decrease as the iteration number $n$ increases. \textbf{b,} Iterative optimization of the estimated ground and excited state energies $\langle E_0\rangle$, $\langle E_1\rangle$, $\langle E_2\rangle$, and $\langle E_3\rangle$ at $R=0.90~\text{\AA}$. Solid colored curves (red, orange, green and blue) denote experimental estimates, and black dashed lines indicate theoretical values. Error bars represent one standard deviation obtained from 1{,}000 Monte Carlo resamplings. \textbf{c,} Estimated energies as a function of interatomic distance $R$. The energy deviations $\Delta E$ (bottom) are defined as the difference between the experimental estimates and the theoretical values.}
    \label{fig3}
    
\end{figure*}

\begin{figure}[!t]
    \centering
    \includegraphics[width=1\columnwidth]{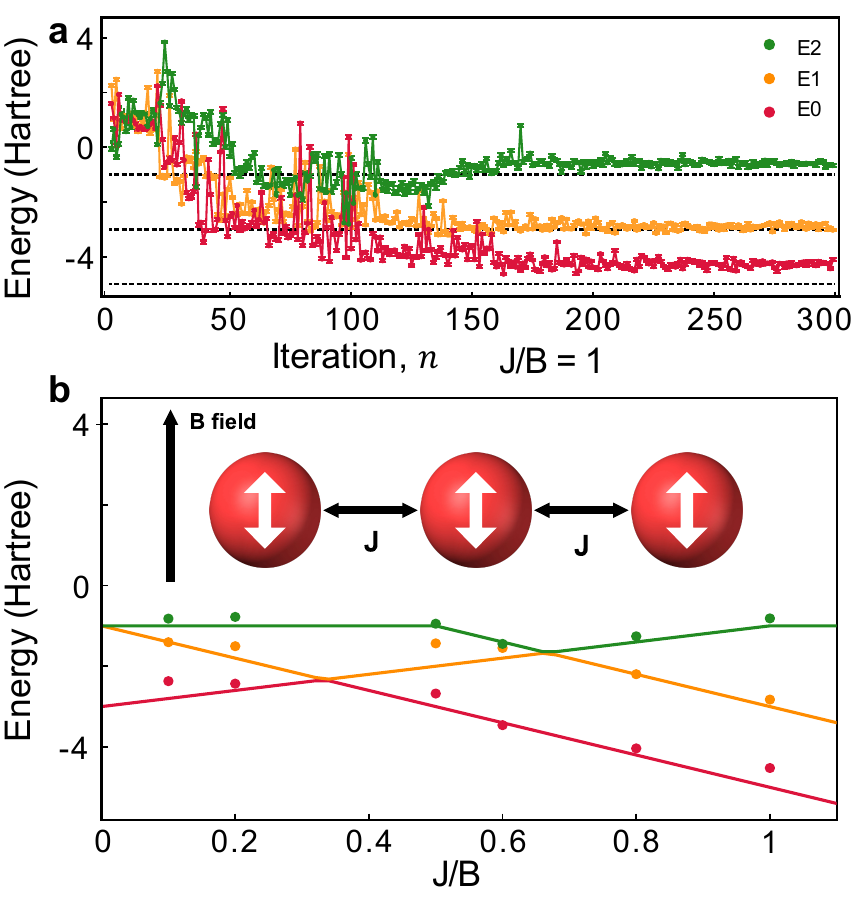}
\caption{\textbf{Experimental SSVQE results for a three-qubit Heisenberg chain.}
\textbf{a,} Iterative optimization of the estimated ground and excited state energies $\langle E_0\rangle$, $\langle E_1\rangle$, and $\langle E_2\rangle$ at $J/B=1$, where the spin--spin interaction and the external field contribute comparably. Solid colored curves (red, orange and green) show experimental estimates as a function of iteration number $n$, and black dashed lines indicate theoretical values. Error bars represent one standard deviation obtained from 1{,}000 Monte Carlo resamplings. \textbf{b,} Estimated energies as a function of the ratio $J/B$.}
    \label{fig4}
    
\end{figure}

\subsection*{Experimental demonstration}

Our experimental setup for SSVQE using OAM qudits is shown in Fig.~\ref{fig2}. A continuous-wave 403 nm laser pumps a 20-mm-long periodically poled potassium titanyl phosphate (PPKTP) nonlinear crystal, generating heralded single photons via type-II spontaneous parametric down-conversion (SPDC)~\cite{harris1967}. A horizontally polarized single-photon is directed onto SLM1 for state preparation. For SSVQE, $N$ mutually orthogonal input states are required, corresponding to the number of energy levels to be estimated. These states are prepared sequentially by displaying the corresponding holograms on SLM1, encoded in an OAM qudit basis~\cite{bolduc2013}. 

For projective measurements, SLM2 displays holograms that implement the measurement bases for the Pauli-string observables. The modulated field is then Fourier transformed in two dimensions and coupled into a single-mode fiber, which projects onto the desired mode and enables estimation of the corresponding expectation values~\cite{bent2015}. The measured outcomes are processed on the CPU to reconstruct the Hamiltonian expectation values and compute the SSVQE cost function $L$, after which an optimizer updates the variational parameters $\vec{\alpha}$. This prepare--measure--update cycle is repeated until convergence. Detailed information on our experimental setup is provided in Supplementary Fig. 1 of Supplementary Note I.

We use both azimuthal index $l$ and radial index $p$ of Laguerre--Gaussian (LG) modes to ensure the orthogonality of the basis while equalizing their coupling efficiencies. For the four-dimensional Hamiltonian, we use $(l,p) \in \{(-3,4),(-1,4),(1,4),(3,4)\}$, and for the eight-dimensional Hamiltonian we use  $(l,p) \in \{(-7,3),(-5,4),(-3,6),(-1,8),(1,8),(3,6),(5,4),(7,3)\}$. The phase maps are provided in Supplementary Fig. 2 of Supplementary Note II. To confirm the orthogonality of the superposed states, we measured the crosstalk matrices for mutually unbiased bases (MUBs). The visibilities of crosstalk matrices and forms of MUBs are provided in Supplementary Figs. 3--5 and Supplementary Tables I and II of Supplementary Note II.

We first benchmark our SSVQE on a four-dimensional $\rm{HeH^+}$ model Hamiltonian to validate the performance of the photonic implementation. Using a single optimization run, we simultaneously estimate the full four-level spectrum of $\rm{HeH^+}$. The Hamiltonian is generated using the OpenFermion--Psi4 library~\cite{mcclean2020}.  $\rm{HeH^+}$ is known to have a minimum energy at an interatomic distance of $R=0.90~\text{\AA}$, corresponding to its equilibrium bond length. Figure~\ref{fig3}{\bf a} shows the optimized ansatz parameters at $R=0.90~\text{\AA}$. We use 15 variational parameters, corresponding to $d^{2}-1$ for $d=4$, and group them into $\vec{\theta}$, $\vec{\phi}$, and $\vec{\omega}$ to conveniently scale the parametrization when targeting a reduced subspace (see Methods and Supplementary Note~III). The parameters are initialized with arbitrary values $\vec{\alpha}_{0}=(\vec{\theta}_{0},\vec{\phi}_{0},\vec{\omega}_{0})$ and converge as the optimization proceeds. As shown in ~\ref{fig3}\textbf{b}, the experimentally obtained energies converge toward the theoretical values (black dashed lines) with increasing iteration number. The weight coefficients $\{w_j\}$ used for the energy estimates are provided in Supplementary Table~III of Supplementary Note~IV.

Figure~\ref{fig3}\textbf{c} summarizes the estimated ground and excited state energies of $\rm{HeH^+}$ as a function of interatomic distance $R$. The solid curves show the theoretical energies, while the markers denote the experimental SSVQE results, revealing good agreement across the full range of $R$ (inset). The lower panel of Fig.~\ref{fig3}{\bf c} plots the energy deviations on a logarithmic scale, highlighting convergence of all four energy levels toward the theoretical values. Using the convergence criterion described above, the optimization terminates after an average of $81.6 \pm 10.2$ iterations.

A key practical feature of SSVQE is that it enables \emph{subspace-selective} eigensolving: only the energy levels of interest need to be targeted, without determining the full spectrum. For example, in a four-level system, SSVQE can be configured to estimate only the ground and first excited states by preparing $N=2$ orthogonal ansatz states. In this case, the number of variational parameters can be reduced accordingly, resulting in a more resource-efficient estimation. Results for the two- and three-level configurations and all iteration graphs for each data are provided in Supplementary Figs. 7--11 of Supplementary Note~V.

To demonstrate scalability beyond the four-dimensional molecular benchmark, we apply our SSVQE to a three-qubit Heisenberg chain, whose Hilbert space is eight-dimensional and can be encoded directly in an OAM qudit. We target the lowest three energy levels within the eight-level spectrum.  The Heisenberg Hamiltonian is given by~\cite{heisenberg1928}
\begin{equation}
H = J \sum_{\langle i,j\rangle} \left(X_iX_j + Y_iY_j + Z_iZ_j\right) + B \sum_{i=1}^{3} Z_i,
\end{equation}
where $J$ is the nearest-neighbor spin--spin interaction strength, $\langle i,j\rangle$ denotes nearest-neighbor pairs $(1,2)$ and $(2,3)$, and $B$ is the external magnetic field along the $Z$ direction. For $J/B \le 1$, the spectrum is dominated by Zeeman splitting with small interaction-induced shifts, whereas for $J/B>1$ the interaction dominates, producing strongly correlated eigenstates and enhanced level repulsion.

Experimentally, we simultaneously estimate three energy levels using an eight-dimensional OAM qudit, with an average convergence of $368 \pm 57$ iterations. Figure~\ref{fig4}{\textbf{a}} shows the iterative convergence of the estimated energies for $J/B=1$. While a general eight-dimensional unitary requires 63 parameters, targeting only three levels allows us to reduce the number of variational parameters to 55 by excluding those associated with unused subspaces (see Supplementary Fig. 6 of Supplementary Note~III for the corresponding parameter evolution).  As shown in Fig.~\ref{fig4}{\textbf{b}}, the estimated energies remain in good agreement with the theoretical values. The remaining experimental deviations are mainly attributed to non-unity purity and state fidelity of the prepared OAM modes, primarily due to imperfect optical alignment. These imperfections are technical rather than fundamental limitations of the scheme and can be further mitigated with improved mode preparation and alignment. Additional results for two-level estimation and all iteration graphs for each data are provided in Supplementary Figs. 12--14 of Supplementary Note~VI.

\section*{Discussion}

In summary, we have demonstrated SSVQE using mutually orthogonal qudit states encoded in the OAM of single photons. By exploiting hologram-based state preparation and measurement on SLMs, our photonic single-qudit platform enables high-dimensional orthogonal states without relying on deep gate sequences. Using this approach, we estimate ground and excited state energies for the $\rm{HeH^+}$ and Heisenberg-chain models in four- and eight-dimensional Hilbert spaces, respectively, and obtain results in good agreement with theoretical values. Our scheme further allows multiple excited states to be estimated simultaneously while scaling the number of variational parameters with the number of targeted energy levels, providing an experimentally practical route to resource-efficient excited state eigensolving.

Our SSVQE implementation offers advantages over many existing excited-state approaches since it does not require ancilla-assisted overlap estimation. In addition, OAM-based qudit encoding provides access to a large (in principle unbounded) photonic Hilbert space~\cite{krenn2014,zeilinger2018,nape2021,bavaresco2018,hu2026}, offering a compact route to high-dimensional quantum simulations with a limited number of photons. 
Nevertheless, this single-qudit encoding has an intrinsic trade-off: encoding an $n$-qubit Hilbert space requires a qudit dimension of $d=2^n$, and the present SLM-based implementation performs projective measurements sequentially by displaying one hologram at a time. Thus, while the present architecture is well suited for near-term demonstrations of high-dimensional variational eigensolvers, scalable photonic quantum computing will ultimately require extensions beyond a single-particle encoding. One promising route is to combine OAM qudits with additional photonic degrees of freedom, including path, polarization, time-bin, and frequency, to further increase the accessible dimensionality ~\cite{barreiro2010,kaneda2019}. Another extension is to use correlated or entangled photon pairs generated via spontaneous parametric down-conversion, where two photons each encoded in a $d$-dimensional OAM space can provide access to a $d^2$-dimensional Hilbert space. 
More generally, extending the scheme to multi-qudit or multi-particle architectures would enable access to larger Hilbert spaces, introducing additional experimental challenges in photon generation, control, and detection~\cite{ringbauer2022,chi2022,wang2022,liu2026,lee2026,kim2026}. We expect our method to be particularly valuable for applications where excited-state properties are central, including fluorescence spectroscopy, battery research, drug discovery, and catalysis~\cite{cao2018}.

\section*{Methods}

\paragraph*{\bf Unitary construction.}
To reduce the number of parameters required to construct high-dimensional unitary operators, we adopt a column-wise synthesis strategy inspired by the triangular mesh architecture of multiport interferometers~\cite{clements2016}. Instead of parameterizing a full $\mathrm{SU}(d)$ transformation with $d^{2}-1$ parameters, we synthesize only the set of $N$ orthonormal columns that define the targeted subspace, and complete the remaining columns by enforcing orthonormality so that the resulting matrix remains unitary. Each column is constructed from a product of two-mode rotations $T_{m}^{(s)}(\theta_s,\phi_s)$, whose block structure enables us to extract only the subset of rotation parameters relevant to the chosen columns. This selective parameterization allows substantial reuse of rotation angles across columns while reducing the dimensionality of the classical optimization.
\begin{align*}
T_{m}^{(s)}(\theta_s,\phi_s) &=
\begin{pmatrix}
1 & 0 &  &  &  &  &  & \\
0 & 1 &  &  &  &  &  & \\
 &  & \ddots &  &  &  &  & \\
 &  &  & e^{i\phi_s}\cos\theta_s & -\sin\theta_s &  &  & \\
 &  &  & e^{i\phi_s}\sin\theta_s & \cos\theta_s &  &  & \\
 &  &  &  &  & \ddots &  & \\
 &  &  &  &  &  & 1 & 0 \\
 &  &  &  &  &  & 0 & 1
\end{pmatrix}.
\end{align*}
To retain generality under this reduction, we prepend a diagonal phase gate $D_{\vec{\omega}}=\mathrm{diag}\!\left(1,e^{i\omega_1},\ldots,e^{i\omega_{d-1}}\right)$, which accounts for relative phase degrees of freedom that would otherwise be absorbed into the unspecified columns. The resulting unitary is written as $U = D_{\vec{\omega}} \prod_{s,m} T_{m}^{(s)}$. For the four-dimensional case, this can be expressed explicitly as $U = D_{\vec{\omega}}\, T_{1}^{(0)} T_{2}^{(1)} T_{0}^{(2)} T_{1}^{(3)} T_{2}^{(4)} T_{0}^{(5)}$. This construction reduces the parameter space while retaining sufficient expressibility within the synthesized subspace.

\section*{Data availability}
The datasets generated and/or analyzed during the current study are not publicly available in accordance with the data management policies of the Korea Institute of Science and Technology (KIST), but are available from the corresponding author on reasonable request.
				
\section*{Code availability}
The code used to generate the figures within this paper and other findings of this study are available from the corresponding author upon request.
				
\section*{Author contributions}
H.-T.L. initiated and led the project. K.-M.H., M.-H.S., and H.-T.L. designed the experimental scheme, and K.-M.H., M.-H.S. performed the experiment. K.-M.H., M.-H.S., H.W.K., and H.-T.L. carried out theoretical calculations. K.-M.H., M.-H.S., and H.-T.L. analyzed the experimental data. Y.-S.K., K.J.L. helped to analyze the data and discuss the results. K.-M.H., H.-T.L. wrote the manuscript.
				
\section*{Competing interests}
The authors declare no competing interests.
				
\section*{Additional information}
				
{\bf Correspondence and requests for materials} should be addressed to Hyang-Tag Lim.

\end{document}